\documentstyle[12pt,psfig]{report}
\newcommand{\be}{\begin{equation}}
\newcommand{\ee}{\end{equation}}
\newcommand{\rf}[1]{(\ref{#1})}
\newcommand{\bra}[1]{\langle{#1}|}
\newcommand{\ket}[1]{|{#1}\rangle}
\newcommand{\lb}{\label}
\newcommand{\thalf}{\mbox{\small $\frac{1}{2}$\normalsize}}
\newcommand{\half}{\frac{1}{2}}
\newcommand{\up}{\uparrow}
\newcommand{\down}{\downarrow}
\begin{document}
\begin{titlepage}
\title{
An introduction to numerical methods \\ in low-dimensional
quantum systems\\
}

\author{Andr\'e Luiz Malvezzi \\
Departamento de F\'\i sica \\ Faculdade de Ci\^encias
  Universidade Estadual Paulista \\ Caixa Postal 473, 17015-970,
  Bauru, SP, Brasil}
\date{\today}

\maketitle
 \end{titlepage}
\tableofcontents



\chapter{Introduction}
\lb{intro}

From a mathematical point of view, quantum mechanics can be regarded
in many respects as an eigenvalue problem. Typically, one has to
to calculate eigenvalue and eigenvectors of Hamiltonians. Very often,
 especially in low temperature problems, the knowledge of the ground
 state and the first few excited states yields considerable insight
 into the physics of a given system. Symmetry properties and the
 known quantum numbers of the desired state can be used to reduce
 the Hilbert space. However, even a large reduction factor will
 eventually be overcome by the exponential growth. This means that
  requirements for memory and running time of relatively small
  systems can be prohibitive, even for the best computers, when a
   full diagonalization of a finite cluster is attempted.
Typically, many degrees of freedom have to be integrated out of the
original problem to make it accessible to present day computer
capacities. 

This notes consist of an introductory course 
to the Lanczos Method and Density Matrix Renormalization Group
Algorithms(DMRG), two among the leading numerical
techniques applied in studies of low-dimensional quantum models.
 Another important numerical technique, the Quantum Monte Carlo, is 
discussed by R. R. dos Santos in his notes for this School.  
The numerical approach allows for a direct and unbiased calculation
of physical properties for finite clusters, from which a phase
diagram can be constructed.   

For those outside the field of one- or two-dimensional
quantum systems it may seem that working on a such low-dimensional
is completely irrelevant to three-dimensional reality. However,
there are in fact may reasons for working in lower-dimensional physics,
in both statistical mechanics as well as other fields.

First, restricting study to one dimension continues to demonstrate
itself as an efficient and effective laboratory for the development
and consideration of new theoretical ideas, many of which are
intended for application to real higher-dimensional.
 Both thermal and
quantum fluctuation effects are larger in lower dimensions.

Secondly, there are in fact real systems which demonstrate a high
degree of low-dimensionality, at least in condensed matter physics.

Thirdly, one-dimensional systems share the possibility with higher-dimensional
systems of undergoing {\em quantum phase transitions} as some parameter 
is varied while holding the temperature fixed at $T=0$. In experimental
situations, this parameter could be pressure or doping, for example,
such as in the case of high temperature superconductors where, upon
the variation of doping, the ground state ranges from antiferromagnetic
to superconducting.

This manuscript is organized in the following way. In Section 2  
the Hamiltonians for the Hubbard, $t-J$ and Heisenberg models are defined. 
This models and their generalizations 
are aimed to describe the microscopic features
of matter. Besides that, they present rich matematical aspects.
In Section 3, the idea of studying the models on clusters of a finite size  
in order to extract their physical properties is discussed. The important
role played by the model symmetries is also examined. 
In Section 4 the Lanczos method is presented and some of its most
popular generalizations discussed.
In Section 5 we start by examining the basic ingredients of the DMRG method 
 and presenting the two kinds of DMRG algorithms. Then, the way measurements
of observables are performed in DMRG is discussed and, finally, the most
important additions to the original method are briefly mentioned.
Section 6 contains some final comments and the acknowledgments.

\chapter{Models}
\lb{models}

In this Section we present some representative quantum 
models. These models 
and their generalizations 
have wide
applicability in statistical mechanics 
as well as in condensed matter physics.

\section{The Hubbard Model}
\lb{Hubb}

A starting point for the microscopic approach to electron motion in crystals
 may be obtained by analyzing the energy levels of the atoms involved.
Lattice models are used, because the atomic structure in the physical systems
determines the possible places at which the electron can be found. If two
 neighboring atoms have overlapping orbitals with very similar energies, 
 the orbitals can hybridize and allow electrons to travel from one atom
 to the other. On the other hand, the repulsion between electrons is very 
strong due to their charge. The simplest approximation for the interaction 
between the electrons is to restrict it to the case when both electrons
 are on the same site (or atom). On-site Coulomb repulsion and 
nearest-neighbour (n.n.) hopping are already the terms of the 
Hubbard model\cite{Anderson,Hubbard}, which can be 
characterized by the Hamiltonian
\be
\lb{HHub}
H_{Hub} = -t \sum_{\langle ij \rangle\sigma} 
\left( c_{i\sigma}^{\dagger}c_{j\sigma} + h.c. \right) +
U \sum_{i} n_{i\uparrow}n_{i\downarrow},
\ee
where $c_{i\sigma}^{\dagger}\ (c_{j\sigma})$ is a creation (annihilation) 
operator for an electron of spin $\sigma=\uparrow,\downarrow$ in a
Wannier orbital at lattice site $i$ and $\langle ij \rangle$ denotes n.n. 
pairs. In the sum, bonds $\langle ij \rangle$ are 
summed over only once each. Here
 $t$ is the matrix element for tunneling from one lattice site to a
 neighbouring one, {\it i.e.} the overlap of n.n. electron 
wavefuctions. The letters $h.c.$ denote the Hermitian conjugate of the
immediately preceding term. The fermionic operators obey the anticommutation 
 relations
\be
\lb{comm1}
\left\{c_{i\alpha},c_{j\beta}^{\dagger}\right\} = 
\delta_{ij}\delta_{\alpha\beta}
\ee
and
\be
\lb{comm2}
\left\{c_{i\alpha},c_{j\beta}\right\} = 0.
\ee
Also, $n_{i\sigma}=c_{i\sigma}^{\dagger}c_{i\sigma}$ is the number
 operator for electrons of spin $\sigma$ at site $i$, and $U$ the
Coulomb repulsion.
 The charge or number operator at site $i$ is
\be
\lb{number}
n_i = \sum_{\sigma} n_{i\sigma} = n_{i\uparrow} + n_{i\downarrow}.
\ee
The states of the model are given by specifying the four possible 
configurations of each site (its level can either be empty, contain
 one electron with either of two spins, or two electrons of
opposite spins) on a lattice made of $L$ sites.

However, the knowledge of the parameters $t$ and $U$ is not enough 
to characterize the system. One also needs to know:
\begin{itemize}
\item The dimensionality of the system, {\it i.e.}, whether it is a
one-dimensional (1D) chain, a two-dimensional (2D) plane or a full 
three-dimensional (3D) system.
\item The geometry of the lattice. The lattice structure can introduce 
frustation, and its symmetry properties greatly influence the behaviour 
of electrons traveling them.
\item The boundary conditions (BC), {\it i.e.}, whether the lattice is 
opened, closed or has some special condition at its surface. 
\item The filling or density,
\be
\lb{density}
n = \frac{1}{L} \sum_{i\sigma} \langle n_{i\sigma} \rangle,
\ee
where $\langle ... \rangle$ denotes a ground state expectation value.
\item The temperature $T$ which is the third energy scale along with
 $t$ and $U$.
\end{itemize}

The first term in Hamiltonian Eq. \rf{HHub} is called hopping
 term and the second one Coulomb term. The hopping term alone 
can be shown to lead to a conventional band spectrum and one-electron 
Bloch levels in which each electron is distributed throughout the
entire crystal (a metal). The Coulomb term alone would favor local magnetic 
moments, since it suppress the possibility of a second electron 
(with oppositely directed spin) at singly occupied sites (an insulator).  
When both terms are present the competition among them brings
 about a transition between the metallic phase and the Mott insulating 
phase\cite{MIT}.

\section{The $t-J$ Model}
\lb{tJ}

If the Hubbard model is considered in the limit where $U/t$ is large, 
the {\em strong coupling limit}, the number of doubly occupied sites is
small. This leads to the derivation of effective models, most
 prominently the $t-J$ model\cite{tJ}. In the $t-J$ model, the 
Hubbarb model complexity is reduced by projecting out the states 
with double occupancy. Using this procedure one gets the 
$t-J$ Hamiltonian
\be
\lb{HtJ}
H_{tJ} = -t \sum_{\langle ij \rangle\sigma}
\left( \tilde{c}_{i\sigma}^{\dagger}\tilde{c}_{j\sigma} + h.c. \right) +
J \sum_{\langle ij \rangle} 
\left( \vec{S}_i \cdot \vec{S}_j - \frac{1}{4}n_{i}n_{j} \right),
\ee
where
\be
\lb{spin}
\vec{S}_i = \sum_{\alpha\beta} c_{i\alpha}^{\dagger} 
\thalf\vec{\sigma}_{\alpha\beta} c_{i\beta}
\ee
is the electron spin operator at site $i$, with 
\be
\lb{Pauli}
\vec{\sigma} = \left[\sigma^x,\sigma^y,\sigma^z\right] =
\left[
\left(
\begin{array}{cc}
0 & 1 \\
1 & 0
\end{array}
\right)
,
\left(
\begin{array}{cc}
0 & -i \\
i & 0
\end{array}
\right)
,
\left(
\begin{array}{cc}
1 & 0 \\
0 & -1 
\end{array}
\right)
\right]
\ee
being the Pauli matrices. The tilde on the $c$-operators in the hopping term
refers to the fact that a creation operator cannot introduce an electron on a
 site where another electron, even of the opposite spin, is already located. 
Formally, one would express this as 
\be
\lb{ctilde}
\tilde{c}_{i\sigma}^{\dagger} = c_{i\sigma}^{\dagger}(1-n_{i,-\sigma})
\ \ \mbox{and, thus,} \ \ 
\tilde{c}_{i\sigma} = (1-n_{i,-\sigma})c_{i\sigma}.
\ee

In the derivation of the $t-J$ model from the Hubbard model, one takes in
account intermediate double occupancies between states which are suppressed
by their high energy but help to lower the kinetic energy by making the
electrons more mobile. In such processes, an electron hops on an already
occupied site, then either electron can hop back to the original site.
Double occupancies are forbidden in the $t-J$ model, but the
physics can be included by adding an additional term. Since the
intermediate processes in the Hubbard model are only possible
 if the electrons on the adjacent sites have opposite spin, this
term can be described by an interaction that favors the
 singlet state compared to the triplet state. This is precisely what
 the second term in Hamiltonian Eq. \rf{HtJ} does. 
To show this, we note that $\vec{S}_i \cdot \vec{S}_j$ 
has two eigenvalues: if the two electrons are in a singlet state,
 the eigenvalue is -3/4, and if the electrons form a triplet, the
energy is 1/4. If the effect of the $\frac{1}{4}n_{i}n_{j}$-term
is included, one gets the following energies from the $J$-term:
\[
\begin{array}{rc}
\mbox{one or both sites unoccupied} & 0 \\
\mbox{spins on both sites forming a triplet state} & 0 \\
\mbox{spins on both sites forming a singlet state} & -J. 
\end{array}
\]

Therefore, the term effectively lowers the energy for states in
which two electrons with opposite spin are on adjacent sites.
From second order perturbation theory of the Hubbard model we 
obtain $J=4t^{2}/U$\cite{tJ}.

\section{The Heisenberg Model}
\lb{Heis}

In the case of half-filling $(n=1)$, when $n_i \equiv 1$ for all sites,
hopping becomes impossible in the $t-J$ model. In addition, the
$\frac{1}{4}n_{i}n_{j}$-term is reduced to a constant. Thus, both terms
can be neglected and the $t-J$ model at half filling is a spin 1/2 
Heisenberg model,
\be
H_{Heis} = J \sum_{\langle ij \rangle}
\left( \vec{S}_i \cdot \vec{S}_j \right).
\ee
It is important to remember that this is an effective model, 
{\it i.e.}, although the spins in the model are localized 
they are meant to describe a system of mobile electrons.
The $\vec{S}_i \cdot \vec{S}_j$ interaction is called a spin
{\em exchange} interaction.

All these models were originally designed to describe physical 
systems. The Heisenberg model was introduced before the Hubbard model
 and is not restricted to just spin 1/2 systems. The restriction in 
the above case resulted from the fact that electrons are spin 1/2 fermions.

A possible generalization for these models is to extend the n.n.
summation to more distant neighbours.

\chapter{The Study of Finite Size Systems}
\lb{Finite}

Having set up Hamiltonians which are believed to contain the 
physics we are interested on, we are left with the formidable 
task of calculating measurable quantities to gain some 
understanding of the models.

Here we choose to work with two unbiased numerical
techniques namely, Lanczos and DMRG.
These methods are unbiased ({\em i.e.}, in contrast to mean-field
based approaches) in the sense that they do not make an initial assumption
on the nature of the ground state of the system. However, as
we will see later, each of these numerical methods has its limitations.

In studying finite size systems the general idea is to construct
a matrix representation of the Hamiltonian for a given
number of sites (system size). 
The Hamiltonian  is then diagonalized in order to obtain
the spectra and calculate measurable quantities, {\em e.g.},
spin and charge correlation functions. We repeat that for
systems of different sizes and extrapolate  the results
toward an infinite size system ({\em i.e.}, towards 
the thermodynamic limit). This agenda is usually well
succeeded provided that we have enough data for a 
reliable extrapolation. The number of lattices with different
sizes needed for the extrapolation procedure to converge depends
heavily on the model being studied and even on the 
set of parameters  being used for a given model. 
The Finite Size Scaling theory\cite{FSS} is specially useful
when critical behaviour is present since, in this case,
a strong dependence of the physical quantities on the system
size is expected\cite{Critical}.  
However, in many cases a fairly good ideia of the properties in the
 thermodynamic limit can be achieved by examining  
just a few system sizes.
The models we are dealing with involve many parameters ({\it e.g.},
$n,\ t,\ U,$ and $J$ in the models of Section \ref{models}). In order
to build up a phase diagram we have to systematically cover the
model parameter space to see how the physical quantities depend
 on each of these parameters.

We can roughly divide the above task in two steps:
\begin{itemize}
\item Build up a representation for the Hamiltonian, diagonalize it, 
and calculate the measurable relevant quantities.
\item Interpret the results and construct the phase diagram.  
\end{itemize}

The second step strongly depends on the model being studied and  
on what kind of physics we are looking at. If the subject
being studied is an advanced topic in physics (as it usually is),
 then a considerable experience in that field of research
might be necessary to carry that step out.  On the other hand,
 the first one is, in principle, much more simple and 
 involves the knowledge of basic concepts in quantum
mechanics and numerical analysis; 
topics covered in most undergraduation courses in physics.
The Lanczos and DMRG are tools for carrying out 
the first step. To understand why such special
techniques are needed let us suppose the
Heisenberg model is to be studied on a lattice of $L$ sites.
Each site has two possible states: spin up and 
down. A lattice with $L$ sites has $2^L$ states
and this is the dimension of the Hamiltonian matrix.  
Similarly, for the $t-J$ and Hubbard models we have 
$3^L$ and $4^L$, respectively.
Due to this exponential growth with $L$ even small lattices
(typically 10 sites or so) generate Hamiltonians too big   
to be handled by present-day computers using standard diagonalization
algorithms.

\section{Symmetries}
\lb{symm}

In the process of constructing a representation for the Hamiltonian 
it is very useful to take advantage of the model symmetries.
Many models, including those presented in Section \ref{models},
 exhibit conservation of total spin number, total
spin in the $z$-direction and total charge, {\em i.e.},
\be
\lb{comm3}
\left[ H,(\vec{S})^{2} \right] = \left[ H,S^z \right] = \left[ H,N \right] = 0,
\ee
where $H$ is the model Hamiltonian and
\begin{eqnarray}
\lb{SN}
\vec{S} & = & \sum_{i} \vec{S}_i \\
N & = & \sum_{i} n_i.
\end{eqnarray}
In addition, these operators also commute with one another, {\em i.e.},
\be
\lb{comm4}
\left[(\vec{S})^{2},S^z\right] = \left[S^z,N\right] =
 \left[N,\vec{S}^2\right] = 0,
\ee
so that the eigenvalues of $H, \vec{S}, S^z$, and $N$ are simultaneous
good quantum numbers, which we will denote simply by $E, S(S+1), S^z$, and
$N$ respectively.\footnote{It will be clear from the context whether
 by $S^z$ or $N$ we mean the operator or its eigenvalue.} 
In the numerical treatment of a given model it is possible to consider 
eigenstates which simultaneously diagonalize $H$ and all operators 
associated to its symmetries. We do this by choosing to work in a
 representation in which the symmetry operators are always diagonal, 
selecting a subspace or sector of Hilbert space
with particular eigenvalues of
those operators, and diagonalizing $H$ in this particular sector.
As exemplified below, total
spin in the $z$-direction and total charge are easily implemented.
However, total spin quantum number $(\vec{S})^2$ is much more difficult 
to specify, and also quite difficult to measure, since in terms of
 the fundamental operators $c_{i\sigma}$ it is extremely non-local:
\be
\lb{S2}
(\vec{S})^2 = \sum_{ij} = \sum_{aij\alpha\beta\gamma\delta}
c_{i\alpha}^{\dagger}
\thalf\sigma_{\alpha\beta}^{a} c_{i\beta}
c_{j\gamma}^{\dagger}
\thalf\sigma_{\gamma\delta}^{a} c_{j\delta},
\ee
where $i$ and $j$ both run throughout the entire lattice, {\em i.e.}, they
do not just represent n.n. terms. 
However, in cases in which choosing the direction of quantization
 in the $z$-axis is arbitrary, {\em i.e.}, in which there is full 
rotational (or $SU(2)$) symmetry, 
the eigenvalues of $H$, $N$, and $\vec{S}^2$ 
will be independent of the eigenvalues of $S^z$, which may assume 
any value ranging from $-S$ to $S$. This can be quite useful in 
determining the total spin quantum number $S$ of the ground state 
if it is only possible to specify the projection $S^z$ in the 
method of investigation.

Let us denote by $E(S^{z})$ the ground state in the sector specified 
by the spin projection $S^z$. If $E(S^{z}) < E(S^{z}+1)$ and 
$E(S^{z}) = E(S^{z}-1) = ... = E(S^{z}_{min})$, where
\be
S^{z}_{min} = \left\{ 
\begin{array}{ll}
0 & \mbox{if $S^z$ integer}\\
\frac{1}{2} & \mbox{if $S^z$ half-odd-integer},
\end{array}
\right.
\ee
then the absolute ground state, ({\em i.e.}, that in the Hilbert
space unrestricted by the specification of $S^z$) contains a 
state of spin $S^{z}$ and no states of any higher spin.

Strongly correlated electron models also often exhibit a {\em particle-hole} 
symmetry. This symmetry relates the creation of an electron to its 
destruction in the following way. Consider the transformation
\be
\lb{PH1}
{\rm PH :\ } c_{i\sigma} \rightarrow (-1)^i c_{i\sigma}^{\dagger}.
\ee
Under this transformation, the n.n. hopping terms transform according to
\be
\lb{PH2}
{\rm PH :\ } c_{i\sigma}^{\dagger}c_{i+1,\sigma} + h.c. \rightarrow 
-c_{i\sigma}c_{i+1,\sigma}^{\dagger} + h.c. = 
c_{i+1,\sigma}^{\dagger}c_{i\sigma} + h.c. =
c_{i\sigma}^{\dagger}c_{i+1,\sigma} + h.c.,
\ee
remaining unchanged under this transformation. However, the number 
operators transform according to
\begin{eqnarray}
\lb{PH3}
{\rm PH :\ } n_{i\sigma} = c_{i\sigma}^{\dagger}c_{i\sigma} 
& \rightarrow & 
c_{i\sigma}c_{i\sigma}^{\dagger} = 
1 - c_{i\sigma}^{\dagger}c_{i\sigma} = 1 - n_{i\sigma} \\
{\rm PH :\ } n_i & \rightarrow & 2 - n_i \\
{\rm PH :\ } N & \rightarrow & 2L - N,
\end{eqnarray}
and similarly the conduction electron spin operator transforms 
according to
\begin{eqnarray}
\lb{PH4}
{\rm PH :\ } \vec{s}_{i} = (s_{i}^{x},s_{i}^{y},s_{i}^{z})
& \rightarrow & 
(-s_{i}^{x},s_{i}^{y},-s_{i}^{z}) = {\rm R}\vec{s}_{i} \\
{\rm PH :\ } \vec{S} = (S^{x},S^{y},S^{z})
& \rightarrow & 
(-S^{x},S^{y},-S^{z}) = {\rm R}\vec{S} \\
{\rm PH :\ } \vec{S}^2 & \rightarrow & \vec{S}^2.
\end{eqnarray}
Here R represents a spin rotation by $\pi$ about the $y$-axis. If this 
particle-hole symmetry applies to the full Hamiltonian, {\em i.e.} not
 only to its non-interacting kinetic part as shown here but also 
its interaction terms, then there will be a one-to-one correspondence 
between eigenstates with quantum number $(E,S,S^{z},N)$ and
$(E,S,-S^{z},2L-N)$. In particular, if we wish to determine the properties 
of a system away from half-filling, we need only do this below
 half-filling and the physics above half-filling will be identical.
 
In order to reduce finite size effects it is common to  work with 
periodic boundary conditions (PBC) to eliminate the lattice surface in given
 a direction.  In this case, the system gains translational invariance
in that direction\footnote{In fact, a cyclic BC is enough to gain
translational invariance.}.
This symmetry express itself through the commutation of $H$ with
the translational operator $T$, which can be defined as
\be
\lb{T}
T|d_{1}d_{2}...d_{L-1}d_{L}\rangle = 
|d_{2}d_{3}...d_{L}d_{1}\rangle,
\ee
where
\be
\lb{state}
|d_{1}d_{2}...d_{L-1}d_{L}\rangle = 
|d_{1}\rangle\otimes |d_{2}\rangle\otimes,...,
\otimes |d_{L-1}\rangle\otimes |d_{L}\rangle 
\ee
and $d_{i}$ is the state at site $i$ ({\em e.g.}, for the Hubbard model
$d_{i}$ can be either one electron with spin up or down, two electrons
with opposite spins, or an empty site). In Eq. \rf{T} we have assumed,
for the sake of simplicity, a one dimensional lattice of size $L$.
In higher dimensions translational invariance in each direction can
be treated separately. The eigenvalues of $T$ are 
$e^{i\frac{2\pi}{L}k}, \ \  k = 0,1,2,...,L-1$, which we will label
by the number $k$. An eigenstate $\phi_{k}$ of $T$ with eigenvalue $k$ is 
given by
\be
\lb{Tphi}
\phi_k = C\left[ 1 + U + U^2 + U^3 + ... + U^{L-1} \right]  
|d_{1}d_{2}...d_{L-1}d_{L}\rangle,
\ee
where $U=e^{i\frac{2\pi}{L}k}T$ and $C$ a normalization constant.
If $\phi_k = 0$ then is not possible to construct a 
state with momentum $k$ from 
$|d_{1}d_{2}...d_{L-1}d_{L}\rangle$.
Note that $S, N, S^{z}$, and $k$ are simultaneous good quantum numbers.

Another symmetry shared by
all the models introduced and Section \ref{models}
and many others is the charge-conjugation symmetry. This symmetry
implies a one to one correspondence
between eigenstates with quantum number $(E,S,S^{z},N)$ and
$(E,S,-S^{z},N)$ and is present if the Hamiltonian commutes
with the parity operator, whose effect is to flip the particle
spin $(|\up\ \down0\ 0\up\rangle \rightarrow |\down\ \up0\ 0\down\rangle)$.
In the presence of charge-conjugation symmetry we can choose to work
only with $S^{z} \leq 0$ or only with $S^{z} \geq 0$. 

Let us consider the $t-J$ model on a chain with four sites
under PBC.
Each site can be in one of the following states: one electron with spin up 
$(|\up\rangle)$ or down
$(|\down\rangle)$, or empty $(|0\rangle)$.
The dimension of total Hilbert space is $3^L = 3^4 = 81$.  
We divide the Hilbert space in sectors labeled by
the quantum numbers $N, S^{z}$, and $k$. Below, we denote each 
sector by $[N,S^{z},k,\mbox{(dimension of the sector)}]$ 
and write down the states for some illustrative cases.

\begin{itemize}
\item $[0,0,0,1]$
\item $[1,0.5,0,1]\ [1,0.5,1,1]\ [1,0.5,2,1]\ [1,0.5,3,1]$
\item $[2,0,0,3]$
\begin{eqnarray*}
\phi^{(1)}_{[2,0,0,3]} & = & 
\frac{1}{2}( |0\ 0\up\ \down\rangle + 
|0\up\ \down0\rangle +
|\up\ \down0\ 0\rangle +
|\down0\ 0\up\ \rangle ) \\
\phi^{(2)}_{[2,0,0,3]} & = & 
\frac{1}{2}( |0\ 0\down\ \up\rangle + 
|0\down\ \up0\rangle +
|\down\ \up0\ 0\rangle +
|\up0\ 0\down\rangle ) \\
\phi^{(3)}_{[2,0,0,3]} & = & 
\frac{1}{2}( |0\up0\down\rangle + 
|\up0\down0\rangle +
|0\down0\up\rangle +
|\down0\up0\rangle )
\end{eqnarray*}
\item $[2,0,1,3]$
\begin{eqnarray*}
\phi^{(1)}_{[2,0,1,3]} & = & 
\frac{1}{2}( |0\ 0\up\ \down\rangle + 
i|0\up\ \down0\rangle -
|\up\ \down0\ 0\rangle - 
i|\down0\ 0\up\ \rangle ) \\
\phi^{(2)}_{[2,0,1,3]} & = & 
\frac{1}{2}( |0\ 0\down\ \up\rangle + 
i|0\down\ \up0\rangle -
|\down\ \up0\ 0\rangle -
i|\up0\ 0\down\rangle ) \\
\phi^{(3)}_{[2,0,1,3]} & = & 
\frac{1}{2}( |0\up0\down\rangle + 
i|\up0\down0\rangle -
|0\down0\up\rangle -
i|\down0\up0\rangle )
\end{eqnarray*}
\item $[2,0,2,3]$
\begin{eqnarray*}
\phi^{(1)}_{[2,0,2,3]} & = & 
\frac{1}{2}( |0\ 0\up\ \down\rangle - 
|0\up\down0\rangle +
|\up\ \down0\ 0\rangle -
|\down0\ 0\up\ \rangle ) \\
\phi^{(2)}_{[2,0,2,3]} & = & 
\frac{1}{2}( |0\ 0\down\ \up\rangle - 
|0\down\ \up0\rangle +
|\down\ \up0\ 0\rangle -
|\up0\ 0\down\rangle ) \\
\phi^{(3)}_{[2,0,2,3]} & = & 
\frac{1}{2}( |0\up0\down\rangle - 
|\up0\down0\rangle +
|0\down0\up\rangle -
|\down0\up0\rangle )
\end{eqnarray*}
\item $[2,0,3,3]$
\begin{eqnarray*}
\phi^{(1)}_{[2,0,3,3]} & = & 
\frac{1}{2}( |0\ 0\up\ \down\rangle - 
i|0\up \down0\rangle -
|\up\ \down0\ 0\rangle +
i|\down0\ 0\up\ \rangle ) \\
\phi^{(2)}_{[2,0,3,3]} & = & 
\frac{1}{2}( |0\ 0\down\ \up\rangle - 
i|0\down\ \up0\rangle -
|\down\ \up0\ 0\rangle +
i|\down0\ 0\up\rangle ) \\
\phi^{(3)}_{[2,0,3,3]} & = & 
\frac{1}{2}( |0\up0\down\rangle - 
i|\up0\down0\rangle -
|0\down0\up\rangle +
i|\down0\up0\rangle )
\end{eqnarray*}
\item $[2,1,0,2]$
\begin{eqnarray*}
\phi^{(1)}_{[2,1,0,2]} & = & 
\frac{1}{2}( |0\ 0\down\ \down\rangle + 
|0\down\ \down0\rangle +
|\down\ \down0\ 0\rangle +
|\down0\ 0\down\ \rangle ) \\
\phi^{(2)}_{[2,1,0,2]} & = & 
\frac{1}{\sqrt{2}}( |0\down0\down\rangle + 
|\down0\down0\rangle ) 
\end{eqnarray*}
\item $[2,1,1,1]$
\[
\phi^{(1)}_{[2,1,1,1]}  =  
\frac{1}{2}( |0\ 0\down\ \down\rangle + 
i|0\down\ \down0\rangle -
|\down\ \down0\ 0\rangle -
i|\down0\ 0\down\ \rangle )
\]
\item $[2,1,2,2]$
\begin{eqnarray*}
\phi^{(1)}_{[2,1,2,2]} & = & 
\frac{1}{2}( |0\ 0\down\ \down\rangle - 
|0\down\ \down0\rangle +
|\down\ \down0\ 0\rangle +
|\down0\ 0\down\ \rangle ) \\
\phi^{(2)}_{[2,1,2,2]} & = & 
\frac{1}{\sqrt{2}}( |0\down0\down\rangle - 
|\down0\down0\rangle ) 
\end{eqnarray*}
\item $[2,1,3,1]$
\[
\phi^{(1)}_{[2,1,3,1]}  =  
\frac{1}{2}( |0\ 0\down\ \down\rangle - 
i|0\down\ \down0\rangle -
|\down\ \down0\ 0\rangle +
i|\down0\ 0\down\ \rangle )
\]
\item $[3,0.5,0,3]\ [3,0.5,1,3]\ [3,0.5,2,3]\ [3,0.5,3,3]$ 
\item $[3,1.5,0,1]\ [3,1.5,1,1]\ [3,1.5,2,1]\ [3,1.5,3,1]$ 
\item $[4,0,0,2]\ [4,0,1,1]\ [4,0,2,2]\ [4,0,3,1]$ 
\item $[4,1,0,1]\ [4,1,1,1]\ [4,1,2,1]\ [4,1,3,1]$ 
\item $[4,2,0,1]$ 
\end{itemize}

The example above give an ideia of how 
helpful symmetry implementation can be. 
Indeed, a further  reduction in the Hilbert space can
be achieved in some cases by considering a lattice reflection
symmetry, namely $[W,H_{tJ}]=0$, where
\be
W|d_{1}d_{2}...d_{L-1}d_{L}\rangle = 
|d_{L}d_{L-1}...d_{2}d_{1}\rangle.
\ee
The eigenvalues of $W$ are: +1 and -1.
For instance, the sector $[2,0,0,3]$ can be broken in two
subspaces, indexed by $W = -1\ ([2,0,0,-1,1])$  and +1 $([2,0,0,1,2])$,
and given by:
\begin{itemize}
\item $[2,0,0,-1,1]$
\[
\phi^{(1)}_{[2,0,0,-1,1]} = 
\frac{1}{\sqrt{2}}(\phi^{(1)}_{[2,0,0,3]} -
\phi^{(2)}_{[2,0,0,3]})
\]
\item $([2,0,0,1,2])$
\begin{eqnarray*}
\phi^{(1)}_{[2,0,0,1,2]} & = & 
\frac{1}{\sqrt{2}}(\phi^{(1)}_{[2,0,0,3]} +
\phi^{(2)}_{[2,0,0,3]}) \\
\phi^{(2)}_{[2,0,0,1,2]} & = & \phi^{(3)}_{[2,0,0,3]}.
\end{eqnarray*}
\end{itemize}



In higher dimensions, besides reflection with respect to 
different axes, rotations by several angles are also
available. Note that these operations might or
might not commute with the translational operator
depending on the value of $k$. For instance, reflection
symmetry can not be employed to break sector $[2,0,1,3]$ in sub-sectors.

By implementing the model symmetries we can significantly 
reduce the computational effort required for diagonalizing
the Hamiltonian. In some cases
we can study lattices large
enough to reveal bulk properties
using standard routines to diagonalize each sector of the Hamiltonian.
For instance, a 4$\times$4  Heisenberg system can be fully
diagonalized if most symmetries discussed 
above are implemented\cite{16Heis}.
Since this approach yields the full spectrum we can construct the
exact partition function for the system and, therefore, obtain
the thermal behaviour exactly at 
arbitrary temperature.\footnote{As discussed below,
this is a commodity we will not have when
 working with Lanczos or DMRG methods.}





 

\chapter{Lanczos Method}

In this Section, an algorithm is presented which allows us to determine
 numerically the ground state and some excited states for Hamiltonian 
operators on finite clusters. The basic idea of the Lanczos 
method\cite{Lanczos1,Lanczos2} is
 that a special basis can be constructed where the Hamiltonian has 
a tridiagonal representation. Once in this form the ground state of 
the matrix can be found easily using standard library subroutines such
as {\em Numerical Recipes} or {\em IMSL}. 

The tridiagonal matrix is constructed iteratively.
First, it is necessary to select an arbitrary normalized 
vector $\ket{\psi_{0}}$
in the Hilbert space of the model being studied. The overlap between 
the actual ground state $\ket{\Psi_{0}}$, and the initial state
$\ket{\psi_{0}}$ should be nonzero. If no {\em a priori} information 
about the ground state is known, this requirement is usually easily 
satisfied by selecting an initial state with {\em randomly} 
 chosen coefficients in the working basis that is being used. If some 
other information of the ground state is known, like its total momentum
 and spin, then it is convenient to initiate the iterations with a 
 state already belonging to the subspace having those quantum numbers
 (and still with random coefficients within this subspace).

After $\ket{\psi_{0}}$ is selected, we define a new vector by applying 
 the Hamiltonian $H$ to the initial state. Subtracting the projection 
over $\ket{\psi_{0}}$, we obtain
\be
\lb{lanc1}
\ket{\psi_{1}} = H\ket{\psi_{0}} - 
\frac{\bra{\psi_{0}}H\ket{\psi_{0}}}
{\langle\psi_{0}|\psi_{0}\rangle}
\ket{\psi_{0}},
\ee
that satisfies $\langle\psi_{0}|\psi_{1}\rangle = 0$. Now, we can construct
 a new state that is is orthogonal to the previous two as,
\be
\lb{lanc2}
\ket{\psi_{2}} = H\ket{\psi_{1}} - 
\frac{\bra{\psi_{1}}H\ket{\psi_{1}}}
{\langle\psi_{1}|\psi_{1}\rangle}
\ket{\psi_{1}} -
\frac{\langle\psi_{1}|\psi_{1}\rangle}
{\langle\psi_{0}|\psi_{0}\rangle}
\ket{\psi_{0}}.
\ee
It can be easily checked that $\langle\psi_{0}|\psi_{2}\rangle = 
\langle\psi_{1}|\psi_{2}\rangle = 0$. The procedure can be generalized
 by defining an orthogonal basis recursively as,
\be
\lb{lanc3}
\ket{\psi_{i+1}} = H\ket{\psi_{i}} - a_{i}\ket{\psi_{i}}
- b_{i}^{2}\ket{\psi_{i-1}},
\ee
where $i = 0, 1, 2, ...,$ and the coefficients are given by
\be
a_{i} = \frac{\bra{\psi_{i}}H\ket{\psi_{i}}}
{\langle\psi_{i}|\psi_{i}\rangle}, \ \ \ 
b_{i}^{2} = \frac{\langle\psi_{i}|\psi_{i}\rangle}
{\langle\psi_{i-1}|\psi_{i-1}\rangle},
\ee
supplemented by $b_{0}=0,\ \ket{\psi_{-1}} = 0$. In this basis, it 
can be shown that the Hamiltonian matrix becomes,
\be
\lb{Hlanc}
H  = \left(
\begin{array}{ccccc}
a_0 & b_1 & 0 & 0 & \ldots \\
b_1 & a_1 & b_2 & 0 & \ldots \\
0 & b_2 & a_2 & b_3 & \ldots \\
0 & 0 & b_3 & a_3 & \ldots \\
\vdots & \vdots& \vdots & \vdots & 
\end{array}
\right),
\ee
{\em i.e.}, it is tridiagonal as expected. Once in this form the matrix 
can be diagonalized using standard library subroutines. However, note
that to diagonalize completely the model being studied on a finite 
cluster a number of iterations equal to the size of the Hilbert space 
(or of the subspace under consideration) is needed. In practice this
would demand a considerable amount of CPU time. However, one of the
 advantages of this technique is that accurate enough information about 
the ground state can be obtained after a small 
number of iterations (typically of the order of $\sim$ 100 or less).
The ideia behind Lanczos method is a systematic improvement of a
given variational state that is used to represent the ground state
\cite{LancProof}. The procedure just described assumes $H$ being 
an hermitian matrix. If that is not the case then a generalized
 algorithm is needed\cite{NHLanc}. 

The eigenvalues of \rf{Hlanc} steadily approach the lowest eigenvalues
of $H$ and its eigenstates are expanded in the Lanczos basis
$\ket{\psi_{i}}$. Each state $\ket{\psi_{i}}$ is represented by a
large set of coefficients, when it is itself expanded in the basis 
selected to carry out the problem ({\em e.g}, the basis used in
example of Section \ref{symm}). Thus, in practice, it is not
convenient to store each one of the $\ket{\psi_{i}}$ vectors
 individually, since such a procedure would demand a memory 
requirement equal to the size of Hilbert space sector multiplied by
the number of Lanczos steps. A simple solution to this problem
consists of running the  Lanczos subroutine {\em twice}. 
For instance, if $\ket{\Psi_{0}} = \sum_{i} f_{i} \ket{\psi_{i}}$, 
then in the first run the coefficients $f_{i}$ are obtained, 
and in the second the vectors $\ket{\psi_{i}}$ are systematically 
reconstructed one by one and used to build up $\ket{\Psi_{0}}$
in the original basis.

A common difficulty with the Lanczos method is that finite 
precision arithmetic causes the $\ket{\psi_{i}}$ to lose
 their orthogonality. A consequence of that is the appearance of
spurious eigenvalues (and eigenstates) in the spectra. 
One fix is to repeatedly reorthogonalize,
which is too costly since, in this case, all $\ket{\psi_{i}}$ have to be
stored. Another is to partially reorthogonalize, and a third option
is to ignore the problem. We can choose the later if only the
extremal (lowest or highest few) eigenvalues are needed, which 
is often the case. A final way of avoiding  the problem is 
to stop the Lanczos whenever the problem starts to appear, calculate
the ground state, and use it as an initial state $\ket{\psi_{0}}$ 
to restart the Lanczos procedure. This resets the basis
but keeps the information from the previous Lanczos running. 
By pushing this ideia to its limit, we can perform always just two
Lanczos steps ({\em i.e.}, work just with $\ket{\psi_{0}}$ and 
$\ket{\psi_{1}}$), diagonalize a 2$\times$2 matrix, and use its
lowest eigenstate as a new $\ket{\psi_{0}}$.
This is the so-called {\em modified Lanczos}\cite{modLanc1,modLanc2}.
The modified Lanczos converge slowly than the original Lanczos but
has the convenience of having the ground state always at hand.  
An even more pedestrian technique is the {\em power method}\cite{power},
 which consists of successively applying the Hamiltonian 
to the initial state until all excited states are filtered out 
and only the ground state remains. This procedure is the slowest 
in speed of convergence, but in simple problems is enough and easy
 to program.

One of the greatest appeals of the Lanczos method is the possibility
of calculating {\em dynamical} 
 properties of a given 
Hamiltonian in finite clusters\cite{DinLanc,Lanczos2}.
The technique permits accurate calculations of energy and momentum 
dependent dynamic correlation functions which are observable in 
scattering experiments, such as Neutron Scattering (spin dynamics) 
and Photoemission Spectroscopy which measures the spectral 
function of the system\cite{Dagotto}.
 
Since Lanczos yields a number of excited states another interesting
possibility is the calculation of finite-temperature quantities.
Examples of successful attempts in this front are 
Refs. \cite{alcaraz} and \cite{prelovsek}. 

The main limitation of Lanczos technique is the size of the clusters
that can be studied. Recently, attempts have been made to reach 
larger cluster. The basic ideia is the following. If $\ket{\phi_{i}}$ is
 a complete basis we are working with, then the ground state can 
be formally represented as
\be
\lb{gsexp}
\ket{\Psi_0} = \sum_i g_i \ket{\phi_{i}}.
\ee
In general, all $\ket{\phi_{i}}$ contribute significatively for
the sum but, in some cases, it might happen that several states 
have very small weight $g_i$. In fact, for most models studied a
very  small percentage of the states in the basis dominate the
sum in Eq. \rf{gsexp}. This suggests that useful results can
still be obtained if a large part of the Hilbert space is
simply discarded. The {\em truncated} Lanczos\cite{truncLanc}
implements this idea by systematically enlarging and reducing
the working Hilbert space in a controlled way and by keeping a
fixed number of states (typically a million or so) in the basis.
The number of states kept represent a very small percentage of
the total Hilbert space but the algorithm is designed to search for
and keep
the dominant states in sum Eq. \rf{gsexp}. It is also important
to choose a basis $\ket{\phi_{i}}$ which uses as much information about
the system as possible. For every physical systems we should use a different
basis depending on physical insight we have about the system. For instance,
if we know that there is a tendency to dimerization then it is convenient to
construct a basis of dimers.
A well chosen basis leads to a better
representation of $\ket{\Psi_0}$ for a given number of states kept in the
truncation procedure.

The efficiency of the truncation technique depends heavily on the
model being studied. In particular, it seems suitable for problems
with gaps in the spectrum (like a spingap)\cite{truncLanc2}.

\chapter{Density Matrix Renormalization Group Algorithms}

\section{Overview}
\lb{DMRG1}

The basic agenda to overcome the system size limitations is to use a basis
 in which the ground state can be represented by only few basis
 states. In other words, a procedure must be found to identify
 or construct the important states and neglect or discard all others
 so that the piece of the Hilbert space one operates on remains small.
 The truncated Lanczos, briefly discussed in the previous Section,
is a possible approach to this agenda. It has the advantage of working with
a basis formed by states that have an intuitive meaning so that the results
can be easily interpreted. In addition, dynamical information can be
obtained without difficulty.
The DMRG technique we are about to discuss is an alternative approach
to that agenda. The innovation of the DMRG is that it does not hold
on to a specific basis, but optimizes the basis it uses in the steps
leading to the calculation of the ground state. A disadvantage of the
 DMRG method is that the basis states chosen by the algorithm are
not intuitive, and the description of the state requires the
measurement of observables. For the measurement process, one needs a
representation of the operators in the current basis. Consequently
each operator that needs to be measured must be stored, and every
time the basis is changed all of them have to be transformed. This
is expensive in time and memory. Another disadvantage is that dynamical
 information cannot be easily obtained.

Historically DMRG has its roots in the renormalization group approach
pioneered by Wilson\cite{Wilson}. 
Considering a real-space blocking version for lattice systems of 
Wilson's original approach, the basic idea is to start with a
small system that can be handled exactly. The system size is then
increased without increasing the size of the Hilbert space until
the desired system size is reached.

Increasing the system without increasing the Hilbert space is typically
done in two steps:
\begin{itemize}
\item The system size is increased, and as a consequence, the Hilbert
space grows at the same time.
\item The Hilbert space is truncated to its original size keeping the
system size constant.
\end{itemize}

To characterize such a {\em renormalization} procedure two basic
questions have to be answered:
\begin{itemize}
\item How is the enlargement done ?
\item Which criterion to apply in the second step to differentiate
between the states we will keep from those we will discard ? 
\end{itemize}

In Wilson's like approach, we start with blocks of a certain (small) size.
 In the first step, two such blocks are linked to form a block which
 is twice as large. The Hamiltonian of this larger block is then
exactly diagonalized and its eigenstates are used as basis states. 
The criterion for keeping states is their 
energy, and only eigenstates whose energy lies below a certain threshold 
are kept. The states which are kept characterize the new block that 
is again linked to an identical block, and the process is iterated.

This approach proved to be very effective for the Kondo model\cite{Wilson}.
However, for other strongly correlated systems like those in 
Section \ref{models} it was not successful\cite{renorm}.
The main reason for this failure lies in choosing the block eigenstates 
as the states to be kept. Since the block was not previously connected
to the rest of the system ( another identical block in the case above )
its eigenstates have inappropriate features at the block ends, making
them a poor choice as a basis to represent the ground state of a 
larger system, formed by putting together two (or more) blocks.
This problem was pinpointed by White and Noack in Ref. \cite{White1} and
an attempt was made to fix it by combining eigenstates from several
different blocks under various BC. 
Let us see how DMRG fix this problem. 

\section{Enlargement and Reduction in the DMRG Procedure}
\lb{DMRG2}

As mentioned in Section \ref{DMRG1}, the two most important 
characteristics of a renormalization procedure are the way 
the system grows and how the decision is made on which 
states are kept in the Hilbert space by the truncation step. 
In this Section, these elements of the DMRG procedure will be
discussed. How these elements are used in the global DMRG 
algorithm will be the subject of the subsequent Sections.
\begin{figure}[ht]
\lb{FigDMRG1}
\centerline{\psfig{figure=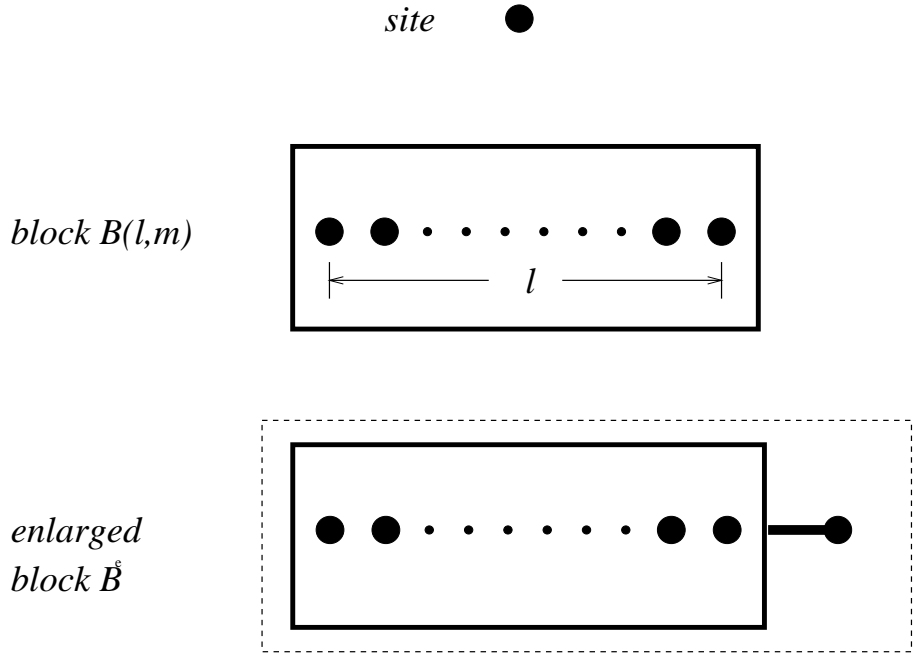,width=12cm,angle=-90}}
\caption{Representation of the basic elements of the DMRG algorithm.
The thick line connecting the block and the site represents all the
interaction terms among them present in the Hamiltonian.}
\end{figure}

Figure 5.1 shows the most important structures used in 
the DMRG algorithm. The elementary unit is a site, and is described 
by the states $d_{i}$ ($i=1,\ldots,D$),
 in which the site can be found\footnote{Here, the index $i$ is not
labeling a site in a lattice 
but the states accessible to a given site. For instance,
$D=4$ and $2$ for the Hubbard and Heisenberg model, respectively.}. 
A {\em block} $B(l,m)$ consists of
a number of sites $l$ and its Hamiltonian $H_{B}$ contains only
terms involving the sites inside the block.
To represent  $B(l,m)$  and $H_{B}$ we associate to them a 
 $m$-dimensional basis where $m$ is in general smaller than the full Hilbert
 space of the block.  The  states in the basis 
are grouped in symmetry sectors labeled by a set of quantum numbers  
({\em e.g.}, $S^{z}$ and $N$), which makes $H_{B}$  a block-diagonal 
matrix. We also store the matrix elements of
$H_{B}$ between these states.  
The block is grown by adding a site to it, and together they form
the {\em enlarged block} $B^e$. If $\ket{b_1}\ldots\ket{b_m}$ 
and $\ket{d_1}\ldots\ket{d_D}$ represent, respectively, the basis of block and a
site then the basis of the enlarged block can be constructed  from the
direct product
\be
\lb{enlarged}
\ket{b^{e}_{k}} = \ket{b_i} \otimes \ket{d_j}.
\ee
The dimension of the Hilbert space for $B^e$ is the product of the
dimensions of the Hilbert space of $B(l,m)$ and a site, {\em i.e},
$m\times D$. A possible mapping of $i$ and $j$-values onto $k$-values 
in Eq. \rf{enlarged} is $k = (i - 1)D + j$.
\begin{figure}[h]
\lb{FigDMRG2}
\centerline{\psfig{figure=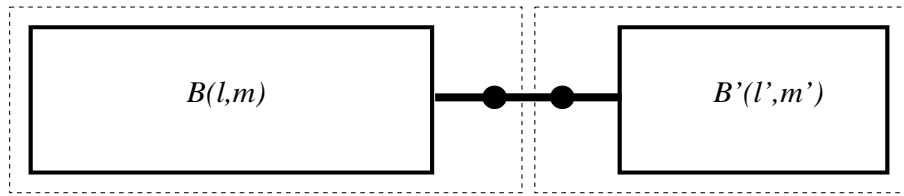,width=12cm,angle=-90}}
\vskip0.2truein
\caption{The superblock consist of two enlarged block connected to
each other. The two sites in the middle are the sites added last
to the respective blocks. In the case of a Heisenberg chain
the two enlarged blocks are connected only by the exchange of
these two sites.}
\end{figure}

The next step in the DMRG method is the formation of the {\em superblock} 
 Hamiltonian (Fig. \ref{FigDMRG2}). The superblock consists of two
enlarged blocks connected to each other. In Fig. \ref{FigDMRG2} open
boundary conditions (OBC) are applied  to the superblock.
This BC are the most widely used in DMRG for it yields the best 
results for a given computational effort. We will discuss PBC latter. 

The DMRG method focus on a single
eigenstate of the superblock Hamiltonian
 (usually the ground state), called the
{\em target state}, which is used to construct the
density matrix\footnote{It is possible to  target several
eigenstates simultaneously but,
for a given computational effort, the accuracy decreases rapidly.}.
The ground state of the superblock is calculated (using Lanczos
or any other method). We then eliminate
the states from the basis of the enlarged block that contribute
the least to the ground state of the superblock. To calculate those,
the density matrix is used.

The concept of the density matrix was developed in statistical
mechanics\cite{Feynman} by considering the problem of a system in
contact with a much larger bath. The ground state of the universe,
{\em i.e.} system and bath, is known, and the question is which states
of the system contribute the most to this ground state. This is what
the density matrix can tell us. One can express the ground state of the
universe (the superblock) in a basis that is the tensor
 product of the basis vectors of the system (one of
 the enlarged blocks) and the bath (the other enlarged block),
\be
\lb{Psi1}
\ket{\Psi_0} = \sum_{i=1}^{m\times D} 
\sum_{j=1}^{m'\times D}
a_{ij} \ket{b_{i}^{e}}\otimes\ket{b_{j}^{'e}}.
\ee
Hence many of the eigenstates of the system contribute to the one
ground state of the universe. The density matrix of the system is given by
\be
\lb{rho}
\rho_{ii'} = \sum_{j=1}^{m'\times D} a_{ij}a_{i'j}^{*}.
\ee
We show an actual example of such a calculation below.
The density matrix has the same dimension and block-diagonal structure
of the Hamiltonian $H_{e}$, for the enlarged block. 
If we denote by $\ket{u_\alpha}$ ($\alpha=1,\ldots,m\times D$) the
eigenstates of $\rho$ and by $w_\alpha$ its eigenvalues then
$\sum_\alpha w_\alpha = 1$ and $w_\alpha$ is the probability of the
 system being in the state $\ket{u_\alpha}$ given that the universe is
in the state $\ket{\Psi_0}$.

This is the information we need to decide which states to keep in a
renormalization group approach. In order to make an optimal decision
of which states to discard and which to keep, it is a good criterion
to consider the weight $w_\alpha$ of the states in the ground state
of a larger system, which we eventually want to describe. 
We must order the $\ket{u_\alpha}$ by their eigenvalues
in a decreasing order and use the $m$ of those states with largest
eigenvalues to form a new basis for the enlarged block $B^e$, which
will then become $B(l+1,m)$. In symbols,
\be
\lb{reduct}
H_{B(l+1,m)} = O\,H_{e}\,O^{\dagger},
\ee
where the rows of the $m\times(m\times D)$ matrix $O$ are formed 
by the $\ket{u_\alpha}$ previously selected. The change of basis
in Eq. \rf{reduct} renormalizes the Hilbert space, cutting its
size back to $m$. Constructed in this way, the blocks are  being 
prepared to be connected to another block in the next step, when
a new superblock will be formed. By using the density matrix
states we somehow {\em look into the future} and adapt the block
for it. Besides $H_B$ we also need to
store other operators representing the sites at the border of the
block. These operators are necessary to construct the interaction
among the block and the site, when forming the enlarged block and
also need to be transformed to according to Eq. \rf{reduct}.
 
To illustrate the DMRG steps of enlargement and truncation, a full
DMRG step for the antiferromagnetic spin 1/2 Heisenberg chain will
be performed\cite{Laukamp}. 
The starting point is a block $B(1,2)$ of a single site.
The possible states of the single site are
\be
\lb{Ex01}
\ket{b_1} = \ket{\up},\ \ \ket{b_2} = \ket{\down}.
\ee
For convenience the up/down basis is chosen.
The basis itself is not stored. The only data that is stored are the
operators needed to progress the algorithm namely, the operators
needed to build the Hamiltonians for the enlarged block and the
superblock.

For one isolated site without external fields the Hamiltonian is zero.
Since the up/down basis was chosen, the other operators are the 
spinmatrices given by
\be
\lb{Ex02}
S^{\pm} = \half \left( \sigma^x \pm i\sigma^y \right),\ \ \
S^z = \half \sigma^z.
\ee
To build the enlarged system, another site is added. In this case 
the basis of the block is the same as the basis of the added site,
\be
\lb{Ex03}
\ket{d_1} = \ket{\up},\ \ \ket{d_2} = \ket{\down},
\ee
and the operators look the same as those of the block. Thus, the
basis of the enlarged block  is, by Eq. \rf{enlarged},
\begin{eqnarray}
\lb{Ex04}
\ket{b_{1}^{e}} & = & \ket{\up \up} \nonumber \\
\ket{b_{2}^{e}} & = & \ket{\up \down} \\
\ket{b_{3}^{e}} & = & \ket{\down \up} \nonumber \\
\ket{b_{4}^{e}} & = & \ket{\down \down}. \nonumber
\end{eqnarray}
The Hamiltonian $H_e$ for the enlarge block $B(2,4)$ 
has non-zero elements, and
describes the interactions of the sites in $B(2,4)$.
$H_e$ consists of the $H_B$, describing the interactions within
the block, and the interactions between the rightmost spin of 
the block and the new site. In the above basis the Heisenberg 
Hamiltonian of the enlarged block is
\begin{eqnarray}
\lb{Ex04a}
H_e & = & H_{B}\otimes I_d + \half \left(S^{+}_{b}\otimes S^{-}_{d} 
+ S^{-}_{b}\otimes S^{+}_{d}\right) + 
S^{z}_{b}\otimes S^{z}_{d} \\
& = & 
\left(
\begin{array}{cc}
0 & 0 \\
0 & 0 
\end{array}
\right) 
\otimes 
\left(
\begin{array}{cc}
1 & 0 \\
0 & 1 
\end{array}
\right) 
\nonumber \\
& & +\half\left[
\left(
\begin{array}{cc}
0 & 1 \\
0 & 0 
\end{array}
\right) 
\otimes 
\left(
\begin{array}{cc}
0 & 0 \\
1 & 0 
\end{array}
\right) 
+
\left(
\begin{array}{cc}
0 & 0 \\
1 & 0 
\end{array}
\right) 
\otimes 
\left(
\begin{array}{cc}
0 & 1 \\
0 & 0 
\end{array}
\right) 
\right] \nonumber \\
& & +\frac{1}{4}
\left(
\begin{array}{cc}
1 & 0 \\
0 & -1 
\end{array}
\right) 
\otimes 
\left(
\begin{array}{rr}
1 & 0 \\
0 & -1 
\end{array}
\right) 
\nonumber
\end{eqnarray} 
and it looks as follows:
 \be
\lb{Ex05}
H_e = \frac{1}{4}
\left(
\begin{array}{rrrr}
1 & 0 & 0 &  0 \\
0 & -1 & 2 &  0 \\
0 & 2 & -1 &  0 \\
0 & 0 & 0 &  1 
\end{array}
\right) .
\ee
In Eq. \rf{Ex04a} the index $b$ and $d$ refers to the operators
acting on the Hilbert space of the block and the site, respectively,
and $I$ is the unit matrix.
In this first step we have $m = D = 2$ but, as the block grows in size in
the following steps we will have  $m > D$.
Note that only representations for the Hamiltonian of the block and
for the operators $S^+,\ S^-$, and $S^z$ of the rightmost site of the
block and the new site are needed to construct the enlarged block.

The superblock is constructed by taking the enlarged block as the left
block and connecting it to another enlarged block on the 
right (Fig. \ref{FigDMRG2}). In the so-called {\em infinite size method},
discussed in the next Section, the right block is the same as the 
left block, only spatially {\em reflected} so that the site last
added to the left block is connected with the site added last to
the right block. The rightmost site of the left block becomes the
leftmost of the right block. 

In addition to the Hamiltonians of the enlarged blocks, one needs
 representations of the spin operators of the rightmost site 
of the enlarged block. In order to construct representations for
$S^+,\ S^-$, and $S^z$ in the basis of the enlarged block we have 
to calculate the tensor product of the unit matrix 
of the block Hilbert space and the operator in the representation 
of the basis of the rightmost site. For instance, the $(S_{r}^{+})_e$ matrix,
the $S^+$-operator of the spin on the $r$ightmost site in the basis 
of the enlarged block, is given by
\be
\lb{Ex06}
(S_{r}^{+})_e = I_b\otimes S^{+}_{d} = 
\left(
\begin{array}{rr}
1 & 0 \\
0 & 1 
\end{array}
\right) 
\otimes 
\left(
\begin{array}{rr}
0 & 1 \\
0 & 0 
\end{array}
\right).
\ee
Representations for $(S_{r}^{-})_e$ and $(S_{r}^{z})_e$ are obtained in 
a similar way. The basis for the superblock is the tensor product of
 the bases from the two enlarged blocks being connected:
\be
\lb{Ex07}
\left(
\begin{array}{l}
\ket{b_{1}^{e}} \\
\ket{b_{2}^{e}} \\
\ket{b_{3}^{e}} \\
\ket{b_{4}^{e}} 
\end{array}
\right)
\otimes
\left(
\begin{array}{l}
\ket{b_{1}^{'e}} \\
\ket{b_{2}^{'e}} \\
\ket{b_{3}^{'e}} \\
\ket{b_{4}^{'e}} 
\end{array}
\right)
=
\left(
\begin{array}{c}
\ket{\up \up} \\
\ket{\up \down} \\
\ket{\down \up} \\
\ket{\down \down} 
\end{array}
\right)
\otimes
\left(
\begin{array}{c}
\ket{\up \up} \\
\ket{\up \down} \\
\ket{\down \up} \\
\ket{\down \down} 
\end{array}
\right)
=
\left(
\begin{array}{c}
\ket{\up \up \up \up} \\
\ket{\up \up \up \down} \\
\ket{\up \up \down \up} \\
\ket{\up \up \down \down} \\
\ket{\up \down \up \up} \\
\ket{\up \down \up \down} \\
\ket{\up \down \down \up} \\
\ket{\up \down \down \down} \\
\ket{\down \up \up \up} \\
\ket{\down \up \up \down} \\
\ket{\down \up \down \up} \\
\ket{\down \up \down \down} \\
\ket{\down \down \up \up}\\ 
\ket{\down \down \up \down}\\ 
\ket{\down \down \down \up}\\ 
\ket{\down \down \down \down} 
\end{array}
\right)
=
\left(
\begin{array}{c}
\ket{b_{1}^{s}} \\
\ket{b_{2}^{s}} \\
\ket{b_{3}^{s}} \\
\ket{b_{4}^{s}} \\
\ket{b_{5}^{s}} \\
\ket{b_{6}^{s}} \\
\ket{b_{7}^{s}} \\
\ket{b_{8}^{s}} \\
\ket{b_{9}^{s}} \\
\ket{b_{10}^{s}} \\
\ket{b_{11}^{s}} \\
\ket{b_{12}^{s}} \\
\ket{b_{13}^{s}} \\
\ket{b_{14}^{s}} \\
\ket{b_{15}^{s}} \\
\ket{b_{16}^{s}} 
\end{array}
\right).
\ee
In general, $\ket{b_{i}^{e}}$ and $\ket{b_{i}^{'e}}$ are distinct basis.
Assuming that we want to calculate the ground state properties, it is
possible to exploit $S^z$ conservation and the fact that the ground state
belongs to the subspace with $S^z = 0$. Therefore, we can concentrate only
on states in this symmetry sector:
\begin{eqnarray}
\lb{Ex08}
\ket{b_{1}^{s(0)}}&  \equiv & \ket{b_{4}^{s}} \nonumber \\
\ket{b_{2}^{s(0)}}&  \equiv & \ket{b_{6}^{s}} \nonumber \\
\ket{b_{3}^{s(0)}}&  \equiv & \ket{b_{7}^{s}} \\
\ket{b_{4}^{s(0)}}&  \equiv & \ket{b_{10}^{s}}\nonumber \\
\ket{b_{5}^{s(0)}}&  \equiv & \ket{b_{11}^{s}} \nonumber \\
\ket{b_{6}^{s(0)}}&  \equiv & \ket{b_{13}^{s}}. \nonumber 
\end{eqnarray}
The Hamiltonian of the superblock consists of three parts: the {\em internal} 
Hamiltonians of the two enlarged blocks and the exchange arising from the
spin interacting at the connection between them:
\be
\lb{Ex09}
H_s = H_e\otimes I_{e}^{'} + I_{e}\otimes H_{e}^{'} + 
\half \left[
(S_{r}^{+})_{e}\otimes (S_{r}^{-})_{e}^{'} +
(S_{r}^{-})_{e}\otimes (S_{r}^{+})_{e}^{'}
\right] + 
(S_{r}^{z})_{e}\otimes (S_{r}^{z})_{e}^{'},
\ee
where the prime refers to operators in the Hilbert space of the 
second enlarged block forming the superblock.   
We now build a representation for $H_s$ in the basis $\ket{b_{i}^{s(0)}}$:
\be
\lb{Ex10}
H_{s}^{(0)} = \frac{1}{4} \left(
\begin{array}{rrrrrr}
1 & 0 & 2 & 0 & 0 & 0 \\
0 & -1 & 2 & 2 & 0 & 0 \\
2 & 2 & -3 & 0 & 2 & 0 \\
0 & 2 & 0 & -3 & 2 & 2 \\
0 & 0 & 2 & 2 & -1 & 0 \\
0 & 0 & 0 & 2 & 0 & 1 
\end{array}
\right).
\ee
The ground state energy of $H_{s}^{0}$ is $E_0 = (1/4)(3+2\sqrt{3})$
and the correspondent eigenvector
\be
\lb{Ex11}
\ket{\Psi_0} = \frac{1}{2\sqrt{3(2+\sqrt{3})}}
\left(
\begin{array}{c}
1 \\
1 + \sqrt(3)\\
-2 - \sqrt{3}\\
-2 - \sqrt{3}\\
1 + \sqrt{3}\\
1 
\end{array}
\right).
\ee

In order to decide which states of the left enlarged block are
the most important for the ground state of the superblock one
uses the density matrix given by Eq. \rf{rho}. Combining 
Eqs. \rf{Ex07} and \rf{Ex08} we obtain
\be
\lb{Ex12}
\left(
\begin{array}{c}
\ket{b_{1}^{s(0)}}\\
\ket{b_{2}^{s(0)}}\\
\ket{b_{3}^{s(0)}}\\
\ket{b_{4}^{s(0)}}\\
\ket{b_{5}^{s(0)}}\\
\ket{b_{6}^{s(0)}}
\end{array}
\right)
=
\left(
\begin{array}{c}
\ket{b_{1}^{e}}\otimes\ket{b_{4}^{'e}}\\
\ket{b_{2}^{e}}\otimes\ket{b_{2}^{'e}}\\ 
\ket{b_{2}^{e}}\otimes\ket{b_{3}^{'e}}\\
\ket{b_{3}^{e}}\otimes\ket{b_{2}^{'e}}\\ 
\ket{b_{3}^{e}}\otimes\ket{b_{3}^{'e}}\\ 
\ket{b_{4}^{e}}\otimes\ket{b_{1}^{'e}} 
\end{array}
\right).
\ee
This allows us to identify the coefficients $a_{ij}$ in Eq. \rf{Psi1}
(and in Eq. \rf{rho}) and they are all zero except for: 
$a_{14},\ a_{22},\ a_{23},\ a_{32},\ a_{33},\ a_{41}$. 
For the density matrix we get
\be
\lb{Ex13}
\rho = \frac{1}{12(2+\sqrt{3})} \left(
\begin{array}{cccc}
1 & 0 & 0 & 0  \\
0 & 11 + 6\sqrt{3} & -2(5 + 3\sqrt{3}) & 0  \\
0 & -2(5 + 3\sqrt{3}) & 11 + 6\sqrt{3} & 0  \\
0 & 0 & 0 & 1 
\end{array}
\right).
\ee
Note that $\rho$ and $H_e$ (Eq. \rf{Ex05})
share the same block-diagonal structure. 
The eigenvalues of $\rho$ are $(1/12)(2 + \sqrt{3}) \approx 0.02$ for
each of the triplet states and $(21 + 12\sqrt{3})(12(2 + \sqrt{3})) 
\approx 0.94$ for the singlet state. The basis states are then
ordered according to the size of the respective eigenvalues, with
 the singlet state (largest eigenvalue) coming first. The transformation
matrix $O$ in Eq. \rf{reduct} is given by
\be
\lb{Ex14}
O = \left(
\begin{array}{cccc}
0 & 1/\sqrt{2} & -1/\sqrt{2} & 0  \\
1 & 0 & 0 & 0 
\end{array}
\right).
\ee
After determining the basis and the transformation, the representations
of all operators used to describe the enlarged block are changed 
to the new basis. Applying the transformation to the $H_e$ (Eq. \rf{Ex05})
leads to
\be
\lb{Ex15}
H_{B(2,2)} = O\,H_{e}\,O^{\dagger} =
\frac{1}{4}\left(
\begin{array}{rr}
-3 & 0    \\
0 & 1  
\end{array}
\right).
\ee
In the present simple case $H_e$ and $\rho$ have the same eigenvectors.
That is the reason why the above transformation diagonalizes $H_e$. 
The same transformation is done with the other operators that will be
need for future calculations. One example is the $S^+$-operator, which
 has the following representation in the new basis
\be
\lb{Ex16}
S_{r}^{+} = O\,(S_{r}^{+})_e\,O^{\dagger} =
\frac{1}{\sqrt{2}}\left(
\begin{array}{rr}
0 & 0   \\
1 & 0 
\end{array}
\right).
\ee
Note that, even though a site has been add to block $B(1,2)$ to form
 block $B(2,2)$, the dimension of Hilbert space did not change due
 to the truncation performed. The states kept in the truncation are
 those with higher probability to be found in the ground states of the
 superblock system.

In this example we have performed the truncation in the Hilbert space
in order to illustrate the procedure. In a practical 
calculation the system in our example would be too small to already
start the truncation. Usually, we know from the beginning how many
states will be kept. Thus, in the first steps the block is grown
(sites are added) without truncation until the number of states needed
to describe the block becomes larger than the number of states we want
 to keep\footnote{In this initial steps with no truncation the matrix $O$
 has dimension $(m\times D)\times(m\times D)$ and Eq. \rf{reduct}
 becomes a simple change of basis.}. If we had, for instance, decide
  to keep $m=20$ states in the computation, the chain would be grown to a
  size of 5 sites with $2^5 = 32$ states, which is the first block size
with Hilbert space dimension exceeding 20.
Then, a truncation would create a block $B(5,20)$ from a enlarged block
with 5 sites and in the following steps all blocks
would have dimension 20, even though they represented a increasing number
of sites.

We often use the sum of the density matrix eigenvalues of the discarded
states ($1 - \sum_{\alpha=1}^{m} w_\alpha$) as a measure for the severity
of the truncation. The goal is to keep this number as small as possible.
In many cases it has been found that this number is roughly proportional
to the error in the energy\cite{Legeza}. The proportionality factor is
of course model dependent. In doped fermionic models, we need to keep
more states to achieve a good accuracy than in a spin model.
Even for a given model the accuracy for a given truncation
may depend on the parameters
being used in the calculation ({\em e.g.}, couplings and symmetry sector).
For instance, close to a phase transition line or inside a critical
(massless) phase the strong quantum fluctuactions tend to reduce accuracy.
In the example above we discarded two of the triplet states leading
 to a truncation error of 0.04, which is unacceptably high. Truncation
  errors in actual calculations are usually kept smaller than $10^{-4}$.

In this Section the focus was on one DMRG step. Enlargement of the block
by adding a site, the formation and diagonalization of the superblock,
the calculation of the density matrix, and the truncation procedure were
discussed. In the next Sections we describe how several DMRG steps are
combined to calculate the properties of a given model.

\section{The Infinite System Algorithm}
\lb{DMRG3}

The first implementation of the DMRG method was the infinite system
algorithm\cite{White2,White3}. The goal was to use DMRG's advantage to
decouple the system size and the size of the Hilbert space and calculate
ground state energies of large systems, {\em i.e.}, system size that are
unreachable for exact diagonalizations, eventually converging to the
thermodynamic limit.
\begin{figure}[ht]
\lb{FigDMRG3}
\centerline{\psfig{figure=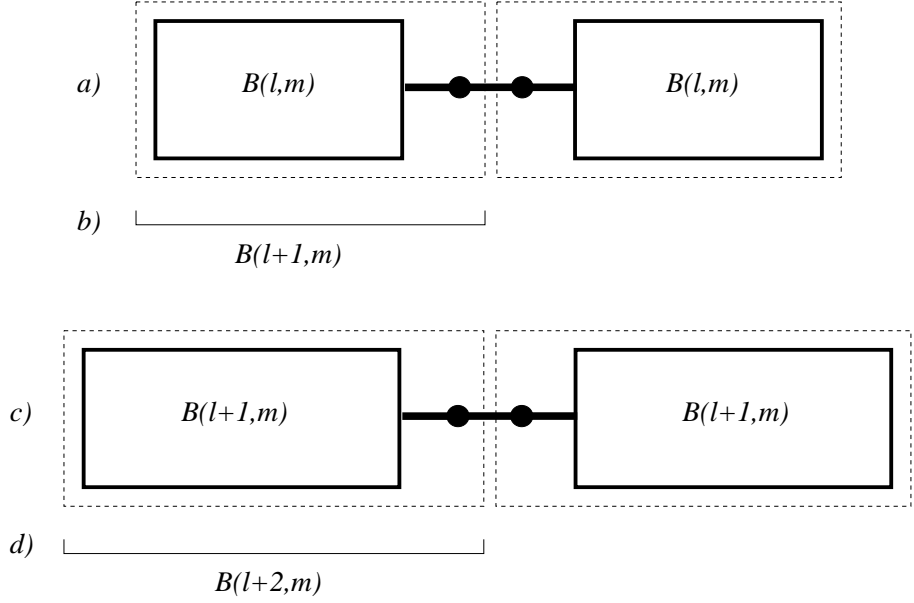,width=12cm,angle=-90}}
\vskip0.2truein
\caption{Two DMRG steps of the infinite system algorithm, see text
for discussion.}
\end{figure}

In the infinite system algorithm, the left enlarged block is connected
by its own mirror image on the right side, so that the number
 of sites of the superblock is increased by two in each step. Growing
 the block and truncating the Hilbert space is done as explained
 in Section \ref{DMRG2}. Schematically the algorithm can be described
 as follows:
\begin{enumerate}
\item Grow the chain to a size in which its Hilbert space dimension
is just larger than $m$, the number of states to be kept. This is the
first enlarged block.
\item Form the superblock by adding an identical enlarged block on the
right such that the sites, which were added last, are next to each other.
\item Diagonalize the superblock, calculate the density matrix with
respect to the left enlarged block.
\item Change basis of the left enlarged block to the eigenbasis of the
density matrix, keep only the $m$ states with the largest density matrix
eigenvalues. The transformed, truncated left enlarged block becomes the
block for the next iteration.
\item This block is enlarged, {\em i.e.}, a site is added on the right.
\item Continue with step 2 until convergence is reached.
\end{enumerate}

In Fig. \ref{FigDMRG3}, two successive iterations of the infinite
system algorithm are shown. The starting point is a block with $l$
sites that is described by a basis with $m$ states. Enlargement and
construction of the superblock (step 2), leads to the situation
portrayed in Fig. \ref{FigDMRG3}.a.

After diagonalizing the superblock, finding the transformation, and
truncating of the enlarged block according to its density matrix
(steps 3 and 4), one arrives at the situation depicted in
Fig. \ref{FigDMRG3}.b. The new block describes a chain with $l + 1$ sites,
but uses a basis with only $m$ states. Enlarging the block (step 6) and
building the superblock (step 2) leads to the situation in
\ref{FigDMRG3}.c, and the procedure is repeated.

From the computational point of view the most difficult part is the
calculation and the subsequent diagonalization of the superblock
Hamiltonian. The diagonalization can be done with the Lanczos method
but any other method ({\em e.g.}, the Davidson algorithm\cite{Davidson})
can be used. The computational effort depends on the size of the
Hamiltonian matrix and accuracy needed for the ground state.
Since the superblock Hamiltonian is block-diagonal, diagonalizing
only the sector that has the proper quantum numbers reduces the matrix
by a factor that depends on the of superblock, block, and model. Typically
it is of order 10-20. Since the total Hilbert space for the superblock
has dimension $(D\times m)^2$, the most important determining factor
 for the size of the Hamiltonian is the number $m$ of states kept in the
 block. In actual calculations $m$ is typically a few hundred of states and
 is limited due to restrictions in computer memory and CPU-time.
As we work with different  models, the size $D$ of the Hilbert space for
a single site also affects the computational effort needed to reach a
given accuracy in the results. 

To summarize: in the infinite system algorithm, the system size is
increased in each step while the number of states kept to describe
the blocks is constant. The goal is to grow the chain to a long-enough
length, so that the energy and short range correlations around the center
have converged. The convergence is checked by keeping track of the
difference $\Delta E_0$ between the ground state energy of the superblocks
in two sucessive steps\footnote{$(\Delta E_0)/2$ converges to the
ground states energy per site in the thermodynamic limit.}.

\section{The Finite System Algorithm}
\lb{DMRG4}

In the finite system algorithm the goal is no longer to reach the
thermodynamic limit, but rather to restrict ourselves to a finite
system size $L$. In  the beginning, until the superblock size reaches
the system size, the algorithm is identical to the infinite size algorithm.
When the system size reaches $L$, {\em i.e.} the enlarged block has
$L/2$ sites, the left block is grown further but on the right a
enlarged block with a smaller number of sites is used in order to
keep constant the number of sites in the superblock. As soon as the
decreasing size of the right block reaches a single site the procedure
is stopped. One such iteration in which the left block has been
calculated for all possible sizes, nearly up to $L$, is called
 a {\em sweep} over the system. After one sweep is done, we start all
 over with a small left block. However, from now on,
 the information about the
 best representation of the right block that complements the left
 block to the desired system size $L$ is now present since it has
 been calculated in the previous sweep. When the optimal basis for
 a specific size of the left block is determined with DMRG, the
 result is stored and used in the next sweep as best possible guess
 for the optimal states describing the right block. The steps of
 the finite system algorithm are compiled in the following list:
\begin{enumerate}
\item In the first sweep use the infinite size algorithm until the
superblock size reaches the chain size $L$ under investigation.
After every truncation save all operators of the reduced block to disk.
\item Enlarge the left block size $l + 1$.
\item Read a block of size $L - l - 2$ from disk, this is the right
block.
\item Enlarge the right block to the size $L - l - 1$.
\item Form the superblock from right and left enlarged blocks.
\item Diagonalize the superblock, calculate the density matrix with
respect to the left enlarged block.
\item Change the basis of the left enlarged block to the eigenbasis of
the density matrix, keep only the $m$ states with the largest
density matrix eigenvalues, save the block with the basis to disk.
The transformed, truncated left enlarged block becomes the left
block for the next step.
\item Continue with step 2 until the right block becomes a single site
\item If the right block is a single site begin a new sweep over the
system {\em i.e.} construct a superblock with a left enlarged block
containing two sites. Continue with step 3 until convergence is reached.
\end{enumerate}

To illustrate the progress of the algorithm, two general steps
are portrayed in Fig. \ref{FigDMRG4}.
\begin{figure}[ht]
\lb{FigDMRG4}
\centerline{\psfig{figure=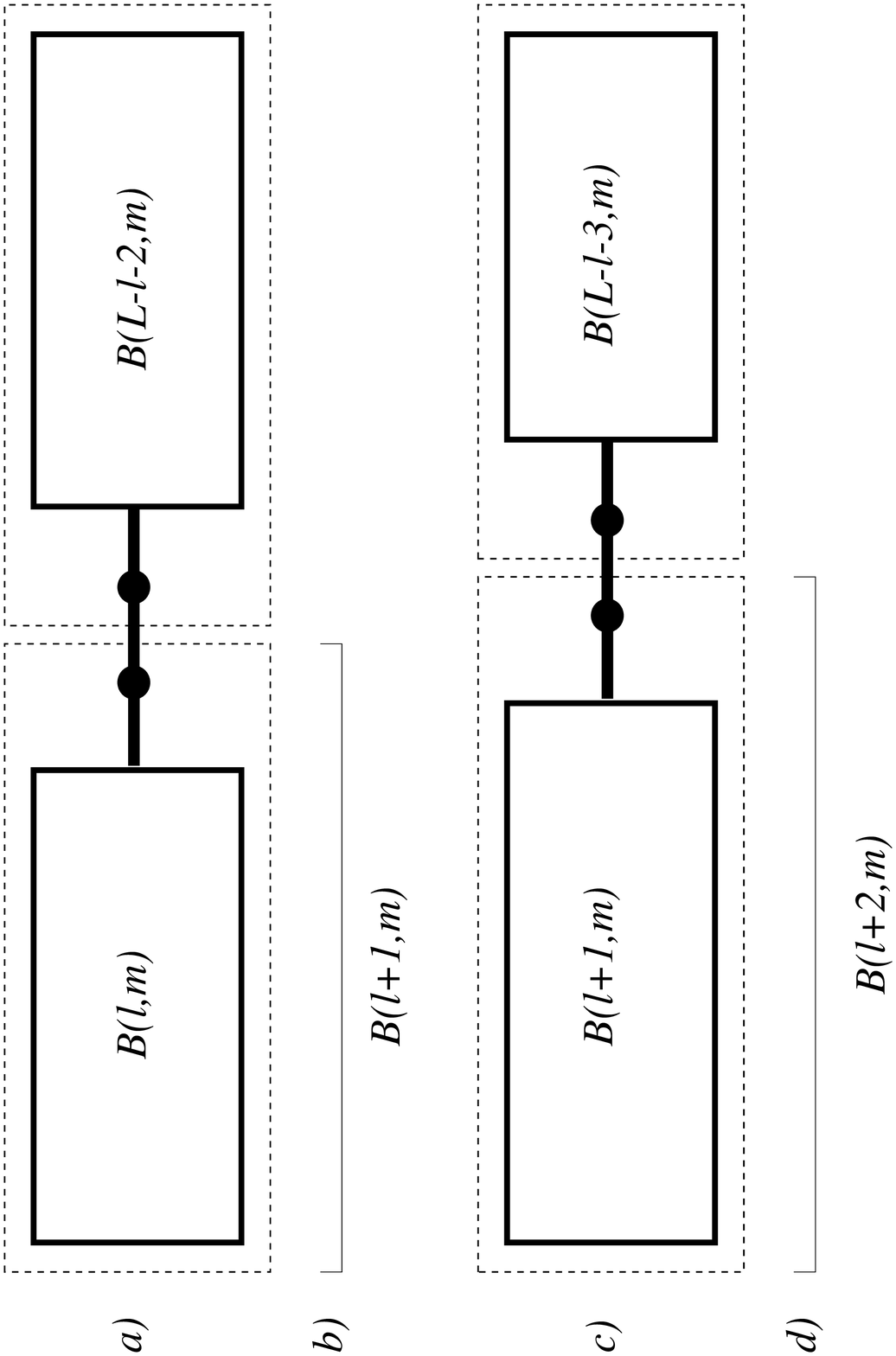,width=12cm,angle=-90}}
\vskip0.2truein
\caption{Two DMRG steps of the finite system algorithm, see text
for dicussion.}
\end{figure}

At first glance Fig. \ref{FigDMRG4} looks like Fig. \ref{FigDMRG3}.
In fact the algorithms are very similar. The left bock is enlarged,
 the superblock is constructed and diagonalized (Fig. \ref{FigDMRG4}.a).
Then the density matrix for the left enlarged block is constructed and
the block is reduced (Fig. \ref{FigDMRG4}.b). This reduced block is
then again enlarged, the superblock is built up and so on. The difference
between the two algorithms is that the right block in Fig. \ref{FigDMRG3}
 has the same number of sites $l$ as the left side. In the finite
 system algorithm the right enlarged block always complements the left one to
 the target size $L$ and thus becomes shorter, while the later
 grows.

As an example, let us consider a calculation for the Heisenberg
chain with $L=16$ sites and a truncation of $m=24$ states.
The following superblocks have to be formed:
\begin{itemize}
\item Initial sweep:

[B(1,2),B(1,2)]\ [B(2,4),B(2,4)]\ [B(3,8),B(3,8)]\ [B(4,16),B(4,16)]

[B(5,24),B(5,24)]\ [B(6,24),B(6,24)]\ [B(7,24),B(7,24)]\ [B(8,24),B(6,24)]

[B(9,24),B(5,24)]\ [B(10,24),B(4,16)]\ [B(11,24),B(3,8)]\ [B(12,24),B(2,4)]
\item Following sweeps:

[B(1,2),B(13,24)]\ [B(2,4),B(12,24)]\ [B(3,8),B(11,24)]

[B(4,16),B(10,24)]\ [B(5,24),B(9,24)]\ [B(6,24),B(8,24)]

[B(7,24),B(7,24)]\ [B(8,24),B(6,24)]\ [B(9,24),B(5,24)]

[B(10,24),B(4,16)]\ [B(11,24),B(3,8)]\ [B(12,24),B(2,4)]
\end{itemize}

If the ground state of the superblock is even under reflection symmetry,
left and right side can be interchanged. That means that the reduced
left blocks stored during the first half of a sweep (when the size of
the left block is smaller than $L/2$) can be already used as right blocks
in the second half of the same sweep. However, if the ground state of
the superblock does not show reflection symmetry, the reduced blocks
from the left and right side have to be constructed and stored independently.
Therefore, in the absence of reflection symmetry stored space and CPU-time
is doubled.

The finite system algorithm is terminated when convergence is reached,
{\em i.e.} when the energy in succeeding sweeps does not improve
(decrease) any more. It happens, however, that the energy stays on a
certain level for two or three sweeps only to further decrease
afterwards. Therefore, one cannot just compare the energies of the last
two sweeps performed. The convergence behaviour of the model should
be taken in account. This behaviour can be investigated by looking at
small systems, where calculations are inexpensive. The number of sweeps
necessary for convergence depends strongly on the system size $L$, the
number of states kept $m$, and the model itself. Fermionic models need
more sweeps than spin models, specially when they are doped. Typically,
spin models converge in less then 10 sweeps, even for fairly
large chains ($L=100$ or larger), while calculations for the doped
$t-J$ model with the same $m$ usually need roughly twice as many
sweeps, even for smaller system sizes.

The first sweeps do not yield very accurate results, their purpose is to
generate a good set of blocks of different sizes. Therefore, the first
sweeps are normally done with a small number of states. When convergence
is approached the number of states kept in the truncation  can be increased
in order to improve accuracy. This helps saving CPU-time, specially when
$L$ is large.

\section{Measurement of Observables}
\lb{DMRG5}

The ground state energy of the superblock is determined every time the
 superblock is diagonalized. The value is used to determine whether
 convergence is reached. It turns out that in the finite system
 algorithm the energy is the lowest when the two blocks forming
 the superblock have the
 same size. Therefore, this {\em symmetric} configuration is used to
 measure also all other observables in which one is interested.

 Unfortunately the values for the other observables such as the
 $S^z$ value of a certain spin or the spin correlation between spins
  on different sites, are not as easily obtainable as the energy.

This is caused by the change of basis that is performed in every step.
Even if we start out with a basis where the demanded properties of the
basis states are known, or could easily be calculated, this  knowledge
 fades fast with the repeated linear combination of basis states of
 new basis systems. Of course we could keep track, for instance, of
  $\langle S^z \rangle$ for each site in each state but the computational
effort would be enormous.

The way that is chosen is to carry out the measurement by acturally
 evaluating the operator in the ground state. The expectation value
 of $S^z$ on site $i$ is calculated as
\be
\lb{Ex17}
\langle S^{z}_{i} \rangle = \bra{\Psi_{0}}S_{i}^{z}\ket{\Psi_{0}}.
\ee
This is the textbook formula, but the difficulty of applying it
here is not visible from this equation. The problem is the basis.
The ground state is expanded in a basis that evolved in every step
of the algorithm and could not have been anticipated at the
beginning.

When the site $i$ is added to the block, the representation of
the $S^z$-operator in the basis of the enlarged block is known
(in Eq. \rf{Ex06} the calculation is done for $S_{i}^{+}$).
But in general this it too early to measure $\langle S^{z}_{i} \rangle$,
 because the symmetric configuration was not yet reached. In order to still
 have the right representation for $S^{z}_{i}$ in the symmetric
 configuration, the matrix has to be updated and stored every time
 the basis changes. Updating means that the basis change has to be
 performed on $S^{z}_{i}$. If $(S^{z}_{i})^{e}_j$ denotes the representation
 of the $S^{z}$-operator on site $i$ in the Hilbert space of the
 enlarged block with $j$ sites ($j\geq i$) and $O_j$ is the matrix that
 transforms and cuts the basis before adding site $j+1$, the representation
 of $S^{z}_{i}$ after the truncation is
\be
\lb{Ex18}
(S^{z}_{i})_{j} = O_j (S^{z}_{i})^{e}_j O^{\dagger}_{j}.
\ee
When another site, the $(j+1)$th, is added, the representation
of the operator also has to be adjusted according to
\be
\lb{Ex19}
(S^{z}_{i})^{e}_{j+1} = (S^{z}_{i})_j\otimes I_{d},
\ee
and truncated with $O_{j+1}$. Following this procedure we always have
representation of the operator in the current basis.

When  it is time to  perform measurements, {\em i.e.} when both blocks forming
the superblock have the same size, we only have to find the
representation of the operator in the Helberb space of the
superblock. If, {\em e.g.}, the operator is acting on a site inside the
left block this means tensorizing it with the unit element acting on
all remaining spaces, namely, the two central sites and right block.

If $i$ is small, {\em i.e.}, the site is close to the end of the
block, a lot of basis changes have to be performed before the
measurement is carried out. Due to the truncations that go with
this procedure, the accuracy is decreased. We expect a greater
accuracy from observables on sites close to the middle of the chain.

The issue is somewhat more complicated for nonlocal operators,
{\em e.g.} spin correlations like $C_{s}(i,j) = \langle S^{z}_{i}
S^{z}_{j}\rangle$. In general one could just take the representations
of the involved $S^{z}$-operators and multiply them, when the symmetric
 configuration is reached. However, there is a more accurate way to
  proceed. As an example, we consider a spin-spin correlation where
   $j = i + 1$ and the symmetric configuration is reached at
$L/2 = i + 2$. The representation of the two operators at the chain
size i+2 are
\begin{eqnarray}
\lb{Ex20}
(S^{z}_{i})_{i+2}^{e} &=& (O_{i+1}((O_i (I_{b}\otimes S^{z})O^{\dagger}_{i})
\otimes I_d )O^{\dagger}_{i+1}\otimes I_b \\
(S^{z}_{i+1})_{i+2}^{e} &=& (O_{i+1}(I_{b}\otimes S^{z})O^{\dagger}_{i+1})
\otimes I_d \nonumber .
\end{eqnarray}
Therefore, one gets for the spin correlation
\be
\lb{Ex21}
(S^{z}_{i}S^{z}_{i+1})_{i+2}^{e} = (O_{i+1}((O_i
(I_{b}\otimes S^{z})O^{\dagger}_{i})
\otimes I_d )O^{\dagger}_{i+1}(O_{i+1}(I_{b}\otimes S^{z})
O^{\dagger}_{i+1})\otimes I_d .
\ee
Another way of calculating the matrix is to multiply the two operators
as soon as possible. In this case the operation can be done when
the enlarged block size is $i + 1$, which gives
\be
\lb{Ex22}
C_{s}(i,i+1))_{i+1} = O_{i+1}
(
(
(
O_i
(
I_b \otimes S^{z}
)
O^{\dagger}_{i}
)
\otimes I_d
)
(
I_b \otimes S^{z}
)
)
O^{\dagger}_{i+1}.
\ee
Then, from now on,   $(C_{s}(i,i+1))_{i+1}$ is tranformed as a whole.
Its representation for the enlarged block with $i+2$ sites is
\be
\lb{Ex23}
(C_{s}(i,i+1))_{i+2}^{e} = C_{s}(i,i+1))_{i+1}\otimes I_d .
\ee
Comparing Eq. \rf{Ex21} with Eq. \rf{Ex23} the difference is only a
a $O^{\dagger}_{i+1}O_{i+1}$-factor between the two $S^z$-operators.
Without the truncation of states it would have been the product
of two unitary matrices and thus be a unit matrix. The two ways
of calculating $(C_{s}(i,i+1))_{i+2}^{e}$ would be equivalent.
With the truncation, however, this is no longer true, as one can
 immediately see calculating $O^{\dagger}O$ with the projector
 in Eq. \rf{Ex14}. The factor $O^{\dagger}_{i+1}O_{i+1}$ leads to
 a loss of accuracy in the matrix multiplication since instead of
  multiplying the matrices, only their projections are multiplied.
This error becomes worse the further the sites $i$ and $j$ in
$C_{s}(i,j)$ are apart, because every separating site add another
pair $O^{\dagger}O$. 

For DMRG procedure this means that we have to use the latter approach,
 multiplying the operators as soon as possible. Prior to the calculation
 a list of the observables that we are interested on measure has to be
 made. When growing the chain, the {\em on-site} operators are stored
 as soon as they are generated and updated every time the basis is
 changed. The products of {\em two-site} operators are formed and
 stored as soon as on has a representation for both operators, then
 they are updated also updated. In the case of operators that involve
  more single site operators, {\em e.g.} pairing operators, we have
  to proceed in the same way.

If we are measuring correlations between sites located on different
blocks we can not multiply them before the symmetric configuration
is reached. This means that measurements of correlations across the
center will always have larger error than correlations where both sites
 are located in the same block.

From these explanations it has hopefully become clear that measuring
observables involves additional storage and calculations. In order to
save computation time the measurement process is started as late as
possible in the calculations, after convergence has been reached.
In the infinite size algorithm we restrict ourselves to the sites
close to the middle of the chain, which were the last to be added.
In the finite size algorithm measurements are not performed in the
initial sweeps when the energy has not converged yet.

It is generally difficult to estimate the error on the values for
the observables other than the energy for which it has been
established that the error is proportional to the truncation error
as discussed before. Unfortunately there is no known method to
calculate the error of the observables from any other measured quantity.
However, by checking how stable the results are as we
change the number $m$ of states kept in the truncation, we can have
a qualitative control of the error.
The error on the energy is generally smaller than that of the observables
because it is a quantity that averages over the sum of many terms.

\section{General Remarks}

When working with fermionic models such as the Hubbard and $t-J$ models
we have to implement the anticommutation of the fermionic operators
Eqs. \rf{comm1} and \rf{comm2}\cite{fermion}. 

In order to implement PBC in DMRG a specific superblock configuration
is required\cite{White3}. DMRG performs poorly under PBC. Typically, if a given
accuracy is obtained under FBC with $m$ states kept in the truncation,
then $m^2$ states will be needed to achieve the same accuracy under
PBC\cite{White3}.

An important improvement to DMRG involves keeping track of the wave 
function from step to step and perform a transformation into the basis
corresponding to the current superblock. Since a good initial guess 
speeds up the Lanczos or Davidson convergence, 
this saves time in the diagonalization of the superblock\cite{White4}.

When DMRG procedure converges to a fixed point the superblock ground state can
be simply written as a matrix-product form and also be rederived 
through a simple variational ansatz making no reference to
the DMRG construction. These very  interesting analytical 
results are obtained in Ref. \cite{Rommer} and give some insight
into the mechanisms working behind DMRG algorithms. 

\chapter{Conclusions and Acknowledgments}

The field of numerical simulations can stand 
by itself as a third way of doing
science and its interaction with experiment and theory is very
fruitful. We believe that scientific research in nearly every area
in physics can benefit from it. 
As the  power of nowadays computers increases rapidly the   
relatively new field of numerical simulations gains more and more prominence.
To keep up with the advances in the hardware new methods and
algorithms are being developed and traditional ones are being improved. 
The two techniques examined here are typical examples of such 
methods and algorithms.

The author is thankful to those (too many to name here!) who carefully read
this manuscript helping to improve it in many ways. The author also
acknowledges hospitality at the Instituto de F\'\i sica de S\~ao Carlos - USP
 and the financial support from Funda\c c\~ao de Amparo \`a Pesquisa do
Estado de S\~ao Paulo - FAPESP - and Conselho Nacional de Desenvolvimento
Cient\'\i fico e Tecnol\'ogico - CNPq - Brasil.  



%

\begin{thebibliography}{02}
%
\bibitem{White1} S. R. White and R. M. Noack,  Phys. Rev. Lett.
{\bf 68}, 3486 (1992).
%
\bibitem{White2} S. R. White, Phys. Rev. Lett. {\bf 69}, 2863 (1992).
%
\bibitem{White3} S. R. White, Phys. Rev. B {\bf 48}, 10345 (1993).
%
\bibitem{Anderson} P. W. Anderson, Phys. Rev. {\bf 115}, 2 (1959).
%
\bibitem{Hubbard} J. Hubbard, Proc. Roy. Soc. {\bf A276}, 238 (1963); 
{\bf A277}, 237 (1964); {\bf A281}, 401 (1964).
%
\bibitem{MIT} For recent review on metal-insulator transitions
see M. Imada, A. Fujimori, and Y. Tokura, Rev. Mod. Phys. {\bf 70},
1039 (1998).
%
\bibitem{tJ} P. W. Anderson, Science {\bf 235}, 1196 (1987); for
the derivation of the $t-J$ model from the Hubbard model see,
{\em e.g.}, P. Fulde, {\em Electron Correlations in Molecules and
 Solids}, Vol. 100 of {\em Solid-State Sciences} (Springer, 1991). 
%
\bibitem{FSS} For a review see, {\em e.g.}, M. N. Barber, Vol. 8 of
 {\em Phase Transitions and Critical Phenomena} (Academic Press,
London, 1984).
%
\bibitem{Critical} S-K Ma, {\em Modern Theory of Critical Phenomena}
(Addison-Wesley Pub., 1997).
%
\bibitem{16Heis} M. Kikuchi and Y. Okabe, J. Phys. Soc. Japan
 {\bf 58}, 679 (1989).
%
\bibitem{Lanczos1} C. Lanczos, J. Res. Nat. Bur. Stand. {\bf 45},
255 (1950).
%
\bibitem{Lanczos2} D. G. Pettifor e D. L. Weaire, {\em The Recursion
 Method and Its Applications}, Springer Series in Solid-State 
Sciences, Vol. 58 (Springer, Berlin/Heidelberg, 1985).
%
\bibitem{Dagotto} E. Dagotto, Rev. Mod. Phys. {\bf 66}, 763 (1994).
%
\bibitem{LancProof} For a more formal discussion of the Lanczos
method see, {\em e.g.}, H. Q. Lin and J. E. Gubernatis,
Computers in Phys. {\bf 7}, 400 (1993) and references therein.
%
\bibitem{NHLanc} See, {\em e.g.}, A. L. Malvezzi, M.A. thesis,
Univ. Federal de S\~ao Carlos - Programa de P\'os-gradua\c c\~ao 
em F\'\i sica (1991).
%
\bibitem{modLanc1} E. Dagotto and A. Moreo, Phys. Rev. D {\bf 31},
 865 (1985).
%
\bibitem{modLanc2} E. R. Gagliano, E. Dagotto, A. Moreo, and
F. C. Alcaraz, Phys. Rev. B {\bf 34}, 1677 (1986).
%
\bibitem{DinLanc} E. R. Gagliano and C. A. Balseiro, Phys. Rev. Lett.
 {\bf 59}, 2999 (1987). 
%
\bibitem{power} A. Jennings and J. J. Mckeown, {\em Matrix Computation},
(John Wiley \& Sons/New York 1983).
%
\bibitem{alcaraz} F. C. Alcaraz and R. R. P. Singh, Phys. Rev. B
{\bf 47}, 8298 (1993).
%
\bibitem{prelovsek} J. Jakli\u{c} and P. Prelov\o{s}ek, Adv. in Phys.
{\bf 49}, 1 (2000).
%
\bibitem{truncLanc} J. Riera and E. Dagotto, Phys. Rev. B {\bf 47},
15346 (1993); J. Pekariwicz and J. R. Shepard, Phys. Rev. B {\bf 56},
 5366 (1997).
%
\bibitem{truncLanc2} E. Dagotto, G. B. Martins, J. Riera, A. L. Malvezzi,
and C. Gazza, Phys. Rev. B {\bf 58}, 12063 (1998). 
%
\bibitem{Wilson} K. Wilson, Rev. Mod. Phys. {\bf 47}, 773 (1975).
%
\bibitem{renorm} J. W. Bray and S. T. Chui, Phys. Rev. B {\bf 19}, 4876
(1979); C. Y. Pan and X. Chen, Phys. Rev. B {\bf 36}, 8600 (1987); 
M. D. Kovarik, Phys. Rev. B {\bf 41}, 6889 (1990); T. Xiang and
 G. A. Gehring, J. Magn. Magn. Mater. {\bf 104-107}, 861 (1992).
%
\bibitem{Feynman} R. P. Feynman, {\em Statistical Physics: A Set of
Lectures} (Benjamim, Reading, MA, 1972).
%
\bibitem{Laukamp} M. Laukamp, {\em Computational Study of ZN-doped 
Quantum Spin Chains and Ladders}, PhD thesis (1998).
%
\bibitem{Legeza} O. Legeza and G. F\'ath, Phys. Rev. B {\bf 53}, 14349
 (1996).
%
\bibitem{Davidson} E. Davidson, Computers in Physics {\bf 7}, 519 (1993);
J. Comput. Phys. {\bf 17}, 87 (1975).
%
\bibitem{fermion} S. Caprara and A. Rosengren, Nuclear Phys. B {\bf B493},
640 (1997); R. M. Noack, S. R. White, and D. J. Scalapino, in {\em Computer
 Simulation Studies in Condensed-Matter Physics VII} - Springer Proceedings
in Physics, Vol. 78 (Edited by D. P. Landau, K. K. Mon, and
 H.-B Sch\"{u}ttler), 85 (1994).
%
\bibitem{White4} S. R. White, Phys. Rev. Lett. {\bf 77}, 3633 (1996).
%
\bibitem{Rommer} S. Rommer and S. \"{O}stlund, Phys. Rev. Lett. {\bf 75},
3537 (1995); Phys. Rev. B {\bf 55}, 2164 (1997). 
%
\end{thebibliography}
\end{document}